\documentclass[epj,twocolumn]{webofc}
\usepackage[varg]{txfonts}   
\usepackage{graphicx}

\wocname{epj}
\woctitle{Seismology of the Sun and the Distant Stars 2016}
\begin{document}
\title{Learning about the latitudinal distribution of starspots through the \\periodogram analysis of photometric data}
%

\author{\firstname{\^{A}ngela R. G.} \lastname{Santos}\inst{1,2,3}\fnsep\thanks{\email{asantos@astro.up.pt}} \and
        \firstname{Margarida S.} \lastname{Cunha}\inst{1,2} \and
        \firstname{Pedro P.} \lastname{Avelino}\inst{1,2} \and
        \firstname{Rafael A.} \lastname{Garc\'{i}a}\inst{4} \and
        \firstname{Savita} \lastname{Mathur}\inst{5}
}

\institute{Instituto de Astrof\'{i}sica e Ci\^{e}ncias do Espa\c{c}o, Universidade do Porto, CAUP, Rua das Estrelas, PT4150-762 Porto, Portugal
              \and Departamento de F\'{i}sica e Astronomia, Faculdade de Ci\^{e}ncias, Universidade do Porto, Rua do Campo Alegre 687, PT4169-007 Porto, Portugal
              \and School of Physics and Astronomy, University of Birmingham, Edgbaston, Birmingham B15 2TT, UK
              \and Laboratoire AIM, CEA/DRF-CNRS-Universit\'{e} Paris 7 Diderot; IRFU/SAp, Centre de Saclay, 91191 Gif-sur-Yvette Cedex, France
              \and Center for Extrasolar Planetary Systems, Space Science Institute, 4750 Walnut street Suite\#205, Boulder CO 80301, USA
          }

\abstract{%
Starspots are cooler and darker than the stellar surface. Therefore, the emitted flux of a star changes when spots are visible on its surface. The presence of spots together with the stellar rotation leads to a periodic modulation on the light curve. By studying that modulation one can then learn about the stellar rotation and also magnetic activity. Recently, Reinhold \& Arlt \cite{Reinhold2015} proposed a method based on the analysis of the Lomb Scargle periodogram of the light curve to identify the sign of the differential rotation, i.e. whether the equator rotates faster than the poles or the opposite. We have been studying in detail the spots’ impact on the light curve and on the resulting periodogram. We find that, under some conditions, the periodogram can actually provide an estimate of the true spot latitudes and/or the stellar inclination angle. Moreover, we find that the impact of the spot on the ratio between the heights of the second and first harmonics of the main peaks in the periodogram can be described by a single parameter, the visibility time of the spot. Finally, we also identify possible sources of false positives/negatives for the sign of the differential rotation.}
\maketitle
%
\section{Introduction}
\label{sec:intro}
The light curves of active stars often show quasi-periodic modulations due to the presence of dark spots crossing the stellar disc. The study of such modulations provides insights on stellar rotation and magnetic properties \cite[e.g][]{Garcia2014}.

The high-precision photometric data, such as those provided by the satellite {\it Kepler} \cite{Borucki2010}, have allowed the detection of the rotation periods for a large number of stars \cite{Nielsen2013}, in particular through the periodogram analysis. 
If the star is differentially rotating, the periodogram of its light curve may show multiple peaks around the detected rotation period. Those peaks are associated to spots at different latitudes and, thus, with different rotation rates. For this reason, one can also use the periodogram analysis to measure the amplitude \cite[e.g.][]{Reinhold2013,Reinhold2013a,Reinhold2015a,Nagel2016,Distefano2016a} and, as proposed more recently, the sign \cite{Reinhold2015} of the differential rotation.

The method proposed by Reinhold \& Arlt \cite{Reinhold2015} relies on the comparison of the peak-height ratios corresponding to different peaks ($P_k$ with $k=1,2,...$) in the periodogram related to the surface rotation. The peak-height ratios $r_k$ are defined as the ratio between the heights of the second and first harmonics of $P_k$, $h_k'$ and $h_k$ respectively, i.e. $r_k=h'_k/h_k$. According to the authors, spots at higher latitudes lead to more sinusoidal light curves than spots at lower latitudes. At these lower latitudes, there is extra power in the second harmonic and, thus, larger ratios $r_k$. Therefore, following their argument, one can attribute a relative latitude ("low" and "high") to each peak $P_k$:

\hspace{0.5cm}\parbox{\hsize-1.cm}{\begin{itemize}
\item[-] ${\rm if}\, r_k>r_{k+1}\Rightarrow P_{\rm low}=P_k\,{\rm and}\,\, P_{\rm high}=P_{k+1}$
\item[-] ${\rm if}\, r_k<r_{k+1}\Rightarrow P_{\rm low}=P_{k+1}\,{\rm and}\,\, P_{\rm high}=P_k$.
\end{itemize}}
Defining the observed relative differential rotation as 
\begin{equation}
\alpha_{\rm obs}=\dfrac{P_{\rm high}-P_{\rm low}}{P_{\rm high}},\label{eq:alpha}
\end{equation}
$\alpha_{\rm obs}$ is positive/negative when the equatorial regions rotate faster/slower than the polar regions (solar/antisolar differential rotation).

In these proceedings, we show the main results from a detailed study of the peak-height ratios and their dependency on the spot and stellar properties (Sect.~\ref{sec:ratios}). We also present the first source for false-positives/negatives of $\alpha_{\rm obs}$ and for observational biases (Sect.~\ref{sec:alpha}).

\section{Results}
\subsection{Peak-height ratios: 1-spot simulations}
\label{sec:ratios}

In this section, we investigate the dependency of the peak-height ratios on a number of stellar and spot parameters which affect the light curve modulation.

Taking the synthetic light curves from 1-spot simulations obtained with a tool that we developed, we start by investigating the impact of the spot latitude and stellar inclination on the peak-height ratios. In this set of synthetic light curves, we fix the limb darkening, spot size and intensity contrast, and the rotation profile. We consider circular spots of radius $R_{\rm S}\sim5.7^\circ$, intensity contrast $C_{\rm S}=0.67$, and infinite lifetime. We also consider solar differential rotation and a quadratic limb-darkening law adequate for solar-like stars.

\begin{figure}[h]
\centering
\includegraphics[trim=5 7 0 0mm,width=\hsize,clip]{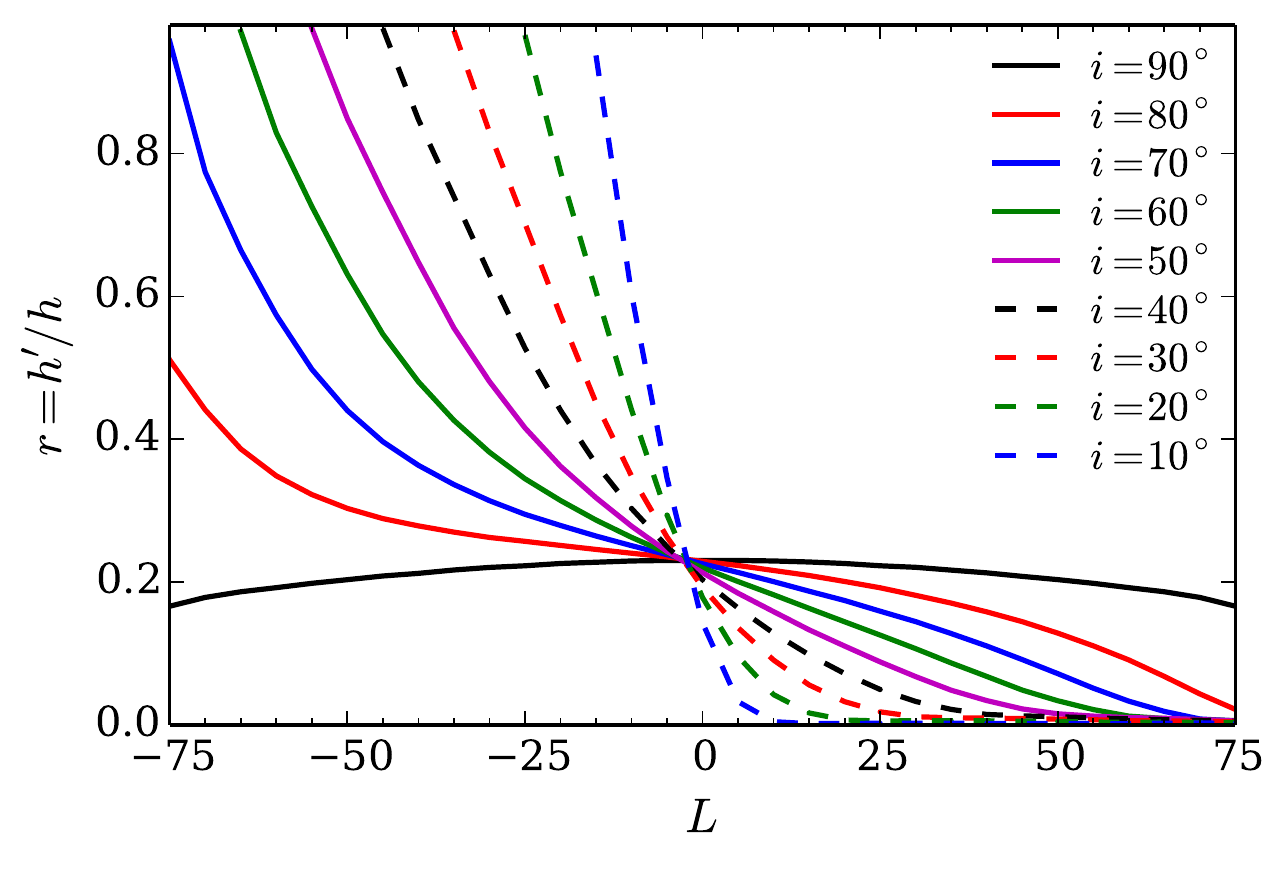}
\caption{peak-height ratios as a function of the spot latitude for different stellar inclination angles.}
\label{fig:ratios}
\end{figure}

For each synthetic light curve, we compute the Lomb Scargle periodogram and, then, the peak-height ratios associated to the rotation period. Figure~\ref{fig:ratios} shows the peak-height ratios as a function of the spot latitude, $L$, for different stellar inclination angles, $i$. From this plot, one can already conclude that the relation between the peak-height ratios and the spot latitude claimed by Reinhold \& Arlt \cite{Reinhold2015} is not fully valid for $i\neq90^\circ$, being only correct for spots lying in the same hemisphere as the observer.

From the results of figure~\ref{fig:ratios}, one can also anticipate the cases where the method proposed by \cite{Reinhold2015} provides the correct sign of the differential rotation:
\begin{itemize}
\item[-] when $i=90^\circ$ (note that the latitudinal dependency of the peak-height ratios for this inclination is weak, which hampers the determination of the sign of the differential rotation);
\item[-] when $i\neq90^\circ$ and the peak-height ratios are related to spots at the northern hemisphere (same as the observer);
\item[-] when $i\neq90^\circ$ and one of the spots is on the southern hemisphere but its absolute latitude is smaller than the latitude of the spots on the northern hemisphere.
\end{itemize}
For the remaining cases, we may expect to recover the wrong sign of the surface differential rotation.

We find that the peak-height ratios plotted against the spot latitudes follow a well defined relation for a given stellar inclination angle. Taking these results as reference for the dependency of the peak-height ratios on the latitude, if the inclination is known, one can estimate the latitude of the spot associated to a given rotation period. Even for the cases where the inclination is unknown, one may constrain the spot latitude and the stellar inclination. 

But what determines the sinusoidality of the spot modulation on the light curve? Are the peak-height ratios a measure of that sinusoidality?

\begin{figure}[h]
\centering
\includegraphics[trim=5 7 0 0mm,width=\hsize,clip]{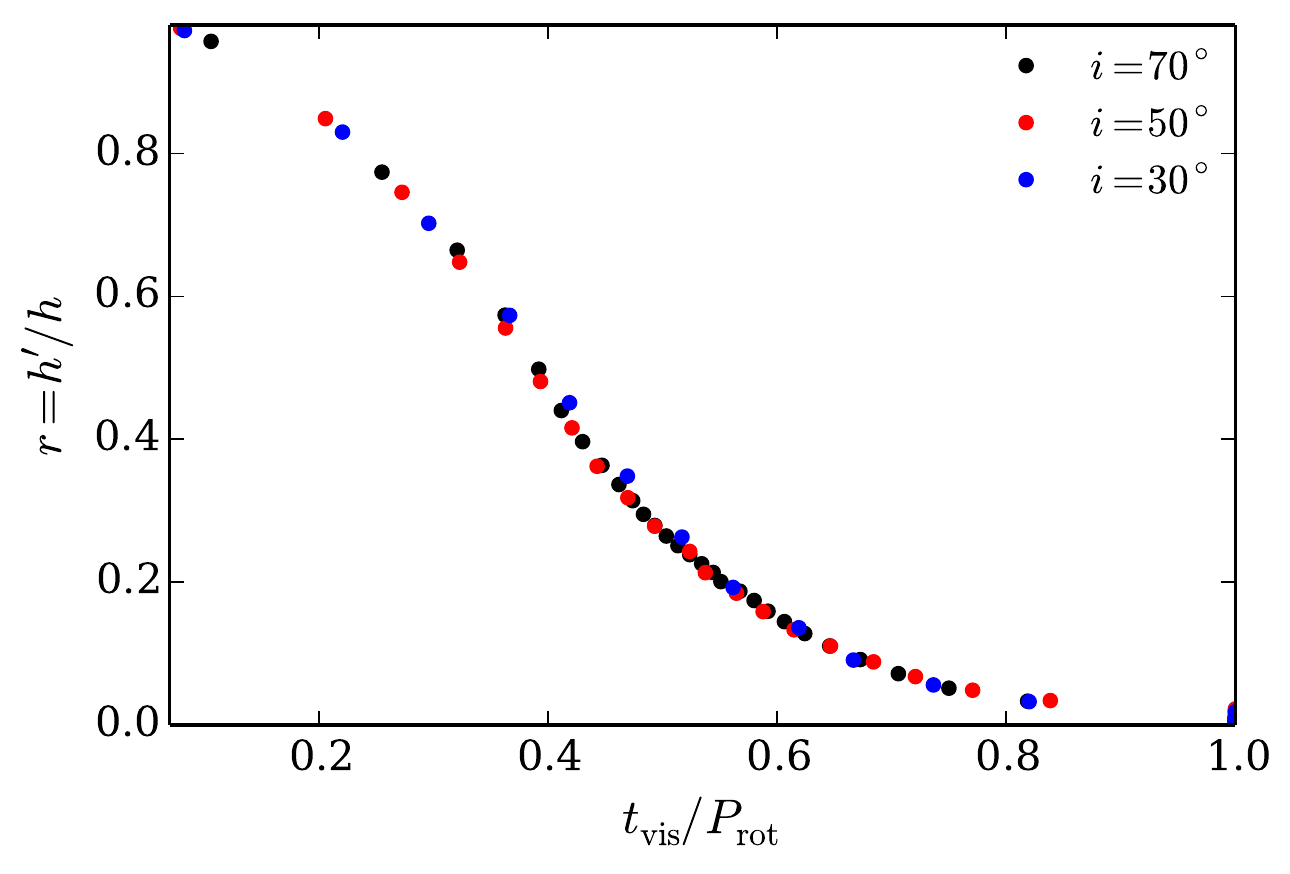}
\caption{peak-height ratios as a function of the visibility time of the spot for different spot latitudes and the stellar inclination angles $i=30^\circ$, $i=50^\circ$, and $i=70^\circ$.}
\label{fig:ratios_tvis}
\end{figure}

We find that the spot latitude and the stellar inclination can be translated into a single parameter related with the visibility of the spot. Figure~\ref{fig:ratios_tvis} shows the peak-height ratios as a function of the ratio between the time the spot remains visible and the rotation period ($t_{\rm vis}/P_{\rm rot}$) for different spot latitudes and stellar inclination angles. The peak-height ratios tend to lie on the same curve, showing that they depend on a single parameter, the visibility time of the spot, and are indeed a measure of the sinusoidality of the light curve.

Besides the stellar inclination and latitudinal distribution of spots, the spot modulation also depends on the stellar rotation and limb-darkening, as well as, on the spot size and intensity contrast. 

In the following simulations, the rotation profile is defined through
\begin{equation}
\Omega(L)=\Omega_{\rm eq}(1-\beta\sin^2L),
\end{equation}
where $\Omega$ and $\Omega_{\rm eq}$ are the angular velocity at a given latitude $L$ and at the equator ($L\!=\!0$), and $\alpha$ is the parameter that determines the amount of surface shear. For the limb-darkening, we consider three different laws: 
\begin{itemize}
\item[-] linear limb-darkening law
\begin{equation}
\dfrac{I(\mu)}{I(1)}=1-u(1-\mu)\end{equation}
\item[-] quadratic limb-darkening law
\begin{equation}
\dfrac{I(\mu)}{I(1)}=1-a(1-\mu)+b(1-\mu)^2\end{equation}
\item[-] 3-parameter non-linear limb-darkening law
\begin{equation}
\dfrac{I(\mu)}{I(1)}=1-c_2(1-\mu)-c_3\left(1-\mu^{3/2}\right)-c_4\left(1-\mu^2\right),
\end{equation}
\end{itemize}
where $I(\mu)/I(1)$ is the relative photospheric intensity, $\mu$ is the cosine of the angle between the line of sight and the normal to a given surface element (at the center of the stellar disc $\mu=1$ and at the limb $\mu=0$), and $u,\,a,\,b,\,c_2,\,c_3$, and $c_4$ are the limb-darkening coefficients, which we take from \cite{Sing2010}.

\begin{figure}[h]
\centering
\includegraphics[width=\hsize,clip]{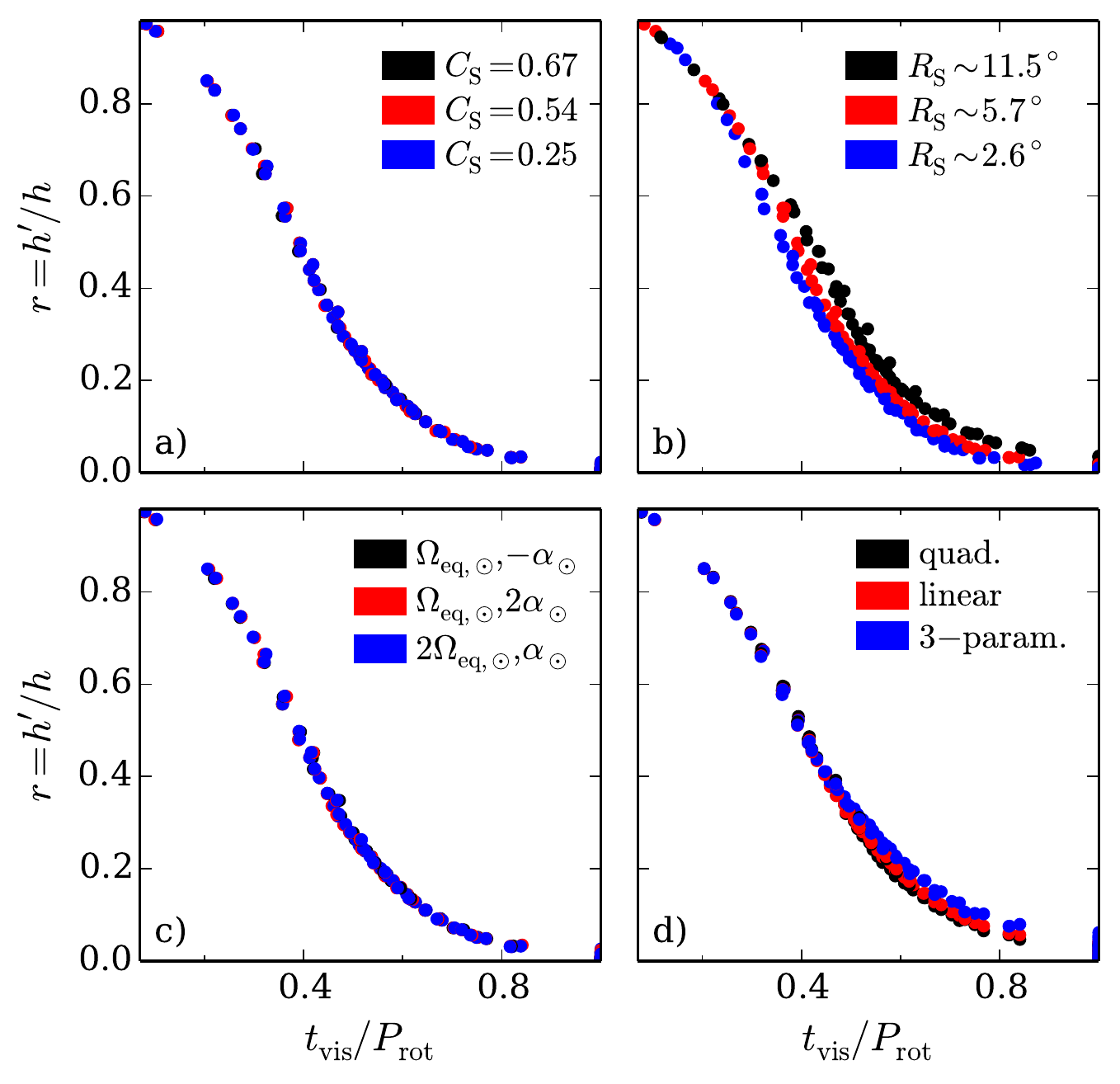}
\caption{peak-height ratios as a function of the spot visibility time for different spot latitudes and stellar inclinations $i=30^\circ$, $i=50^\circ$, and $i=70^\circ$, and different: a) spot intensity contrast, $C_{\rm S}$; b) spot radii, $R_{\rm S}$; c) stellar rotation profiles; and d) limb-darkening laws.}
\label{fig:ratios_par}
\end{figure}

By varying one of those parameters (rotation, limb-darkening, spot size and intensity contrast) at each set of simulations shown in figure~\ref{fig:ratios_par}, we find that the peak-height ratios are independent on stellar rotation and spot contrast (left panels of figure~\ref{fig:ratios_par}), as these parameters do not change the visibility time of the spot, measured as $t_{\rm vis}/P_{\rm rot}$, or the sinusoidality of the modulation. Although the effect is small, we find that the peak-height ratios depend on the limb-darkening (bottom right panel), due to its impact on the shape and sinusoidality of the spot modulation. Finally, the larger the spot the longer it remains visible, thus the spot size affects both peak-height ratios and visibility time of the spot (top right panel). \\

\subsection{Sign of the surface differential rotation: 2-spot simulations}\label{sec:alpha}

This section focuses on the sign of the stellar surface differential rotation. In particular, we present one of the sources for false positives/negatives. 

For each synthetic light curve obtained from 2-spot simulations, we compute the Lomb Scargle periodogram and the corresponding peak-height ratios. In order to avoid the contamination by nearby peaks, we impose an adequate criteria for the peak separation in the periodogram and only consider the peaks that obey that criteria.

\begin{figure*}[t]
\centering
\includegraphics[width=\hsize,clip]{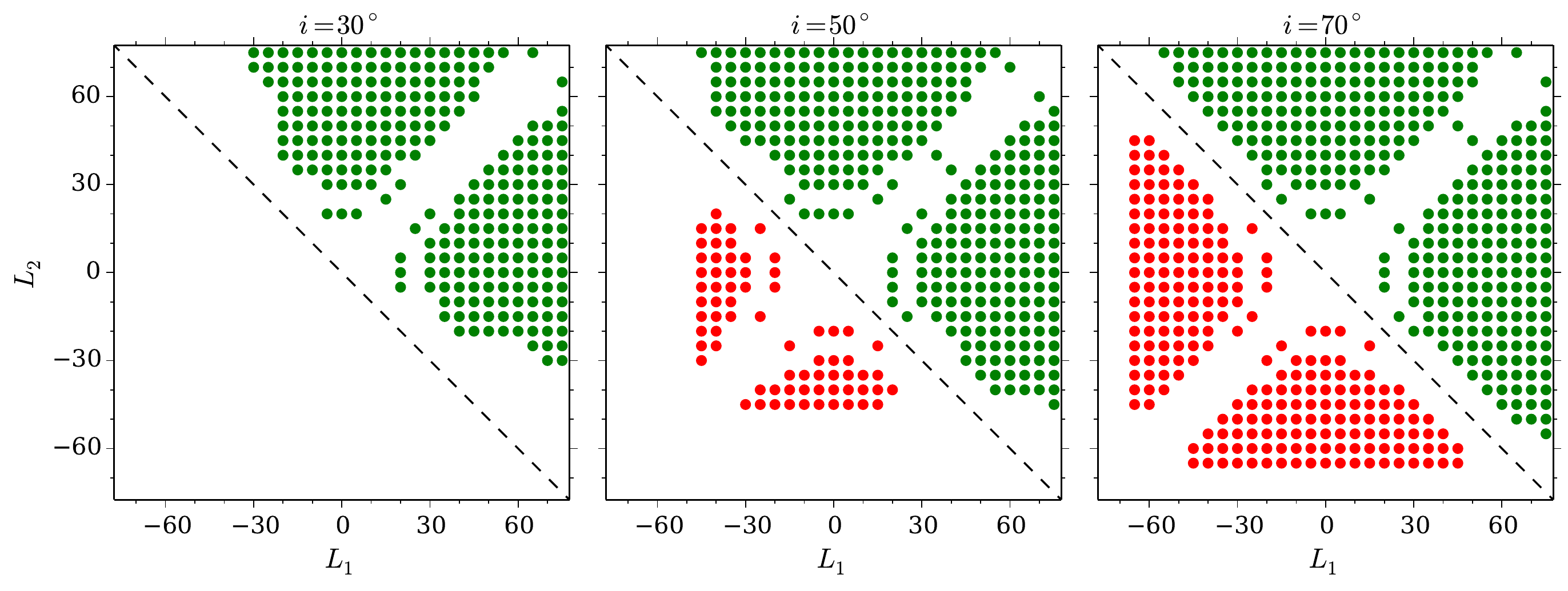}
\caption{Sign of the differential rotation for different 2-spot configurations and inclination angles $i=30^\circ$ (left), $i=50^\circ$ (middle), and $i=70^\circ$ (right). $L_1$ and $L_2$ are the latitudes of each spot respectively. The red dots correspond to the cases of false negatives ($\alpha_{\rm obs}<0$), while the green dots represent the cases where solar differential rotation ($\alpha_{\rm obs}>0$) is correctly recovered. The spot configurations that are not marked in the plots correspond to the cases where no rotation periods are recovered or to the cases where the periods do not fulfill the peak separation criteria. The dashed line divides the regions where the correct (above) or wrong (bellow) sign of $\alpha_{\rm obs}$ was expected from the results presented in figure~\ref{fig:ratios}.}
\label{fig:alpha}
\end{figure*}

Having the peak-height ratios and the rotation periods, we apply the method proposed by \cite{Reinhold2015} and determine the observed relative differential rotation as defined in Eq. (\ref{eq:alpha}). Figure~\ref{fig:alpha} shows the results for the sign of the differential rotation for different spot configurations (from a broad range of latitudes) and stellar inclination angles. The white regions in figure~\ref{fig:alpha} correspond to the spot configurations where the rotation periods are not successfully recovered/distinguishable and, thus, no information on $\alpha_{\rm obs}$ is retrieved. Knowing the input differential rotation ($\alpha>0$, solar differential rotation), the red dots show the cases where the wrong sign of $\alpha_{\rm obs}$ (false-negative) is determined, while the green dots correspond to the cases where solar differential rotation is correctly recovered. We note that we have considered steps of $5^\circ$ for the spot latitudes and spots of equal size in these simulations.

Finally, for small stellar inclination angles, it is more likely to detect spot modulations on the light curve induced by spots at the same hemisphere as the observer than by spots at the opposite hemisphere. Therefore, as shown in figure~\ref{fig:alpha}, the rate of false negatives decreases as we study stars with lower inclinations.

\section{Conclusions}
\label{sec:con}

The main goal of the work presented in these proceedings was to understand to what extent one can retrieve, from the light curve and consequent periodogram analysis, information about the latitudinal distribution of spots and stellar surface differential rotation.

We studied the dependency of the peak-height ratios on the spot and stellar properties. We found that those ratios are essentially a function of a single parameter, the spot visibility time, which is mainly determined by the stellar inclination and spot latitude.

We conclude that, despite the degeneracy between latitude and inclination, the periodogram analysis provides a simple and fast way to constrain these parameters. And if the inclination of the star is known, one can actually estimate the spot latitudes.

Our results also show that, for $i\neq90^\circ$ the method proposed by \cite{Reinhold2015} only returns the correct sign of the differential rotation when:
\begin{itemize}
\item[-] the rotation periods $P_k$ and $P_{k+1}$ are associated to spots at $L_k,\, L_{k+1}>0$ (same hemisphere as the observer);
\item[-] the rotation peridos $P_k$ and $P_{k+1}$ are associated to spots at $L_k>0$ and $L_{k+1}<0$ (same and opposite hemispheres as the observer, respectively) and $|L_k|>|L_{k+1}|$.
\end{itemize}

Finally, we note that the modulation induced by spots at $L_k>0$ will be preferentially observed, specially for small $i$. This will contribute for a low rate of false positives/negatives for the sign of $\alpha_{\rm obs}$.

For further details on this study, we recommend the reader to see \cite{Santos2016a}.\\

\begin{acknowledgement}
\hspace{-0.25cm} This work was supported by Funda\c{c}\~{a}o para a Ci\^{e}ncia e a Tecnologia (FCT) through the research grant UID/FIS/04434/2013. ARGS acknowledges the support from FCT through the Fellowship SFRH/BD/88032/2012 and from the University of Birmingham. MSC and PPA acknowledge support from FCT through the Investigador FCT Contracts No. IF/00894/2012 and IF/00863/2012 and POPH/FSE (EC) by FEDER funding through the programme Programa Operacional de Factores de Competitividade (COMPETE). RAG acknowledges the support of the GOLF and PLATO CNES grants. SM would like to acknowledge support from NASA grants NNX12AE17G and NNX15AF13G and NSF grant AST-1411685. The research leading to these results has received funding from EC, under FP7, through the grant agreement FP7-SPACE-2012-312844 and PIRSES-GA-2010-269194. ARGS, MSC, and PPA are grateful for the support from the High Altitude Observatory (NCAR/UCAR), where part of the current work was developed.
\end{acknowledgement}

%
\bibliographystyle{woc}
\bibliography{diffrot}

\end{document}